\documentclass{sigchi}

\toappear{\scriptsize Permission to make digital or hard copies of part or all of this work for personal or classroom use is granted without fee provided that copies are not made or distributed for profit or commercial advantage and that copies bear this notice and the full citation on the first page. Copyrights for third-party components of this work must be honored. For all other uses, contact the owner/author(s). \\
{\emph{L@S '21, June 22--25, 2021, Virtual Event, Germany.} } \\
\copyright~2020 Copyright is held by the author/owner(s). \\
ACM ISBN 978-1-4503-8215-1/21/06. \\
http://dx.doi.org/10.1145/3430895.3460139}
\usepackage{balance}       % to better equalize the last page
\usepackage{graphics}      % for EPS, load graphicx instead 
\usepackage[T1]{fontenc}   % for umlauts and other diaeresis
\usepackage{txfonts}
\usepackage{mathptmx}
\usepackage[pdflang={en-US},pdftex]{hyperref}
\usepackage{color}
\usepackage{booktabs}
\usepackage{textcomp}

\usepackage{microtype}        % Improved Tracking and Kerning
\usepackage{ccicons}          % Cite your images correctly!
\usepackage{todonotes}

\def\plaintitle{Should College Dropout Prediction Models Include Protected Attributes?}
\def\plainauthor{Renzhe Yu, Hansol Lee, Ren\'e Kizilcec}

\def\plainkeywords{Dropout prediction; Predictive analytics; Higher education; Online learning; Algorithmic fairness}

\makeatletter
\def\url@leostyle{%
  \@ifundefined{selectfont}{
    \def\UrlFont{\sf}
  }{
    \def\UrlFont{\small\bf\ttfamily}
  }}
\makeatother
\def\pprw{8.5in}
\def\pprh{11in}

\definecolor{linkColor}{RGB}{6,125,233}
\hypersetup{%
  pdftitle={\plaintitle},
  pdfauthor={\plainauthor},
  pdfkeywords={\plainkeywords},
  pdfdisplaydoctitle=true, % For Accessibility
  bookmarksnumbered,
  pdfstartview={FitH},
  colorlinks,
  citecolor=black,
  filecolor=black,
  linkcolor=black,
  urlcolor=linkColor,
  breaklinks=true,
  hypertexnames=false
}

\setcopyright{rightsretained}
\begin{document}

\title{\plaintitle}

\numberofauthors{3}
\author{%
  \alignauthor{Renzhe Yu\\
    \affaddr{University of California, Irvine}\\
    \affaddr{Irvine, CA, USA}\\
    \email{renzhey@uci.edu}}\\
  \alignauthor{Hansol Lee\\
    \affaddr{Cornell University}\\
    \affaddr{Ithaca, NY, USA}\\
    \email{hl838@cornell.edu}}\\
  \alignauthor{Ren\'e F. Kizilcec\\
    \affaddr{Cornell University}\\
    \affaddr{Ithaca, NY, USA}\\
    \email{kizilcec@cornell.edu}}\\
}

% \copyrightyear{2021}
% \acmYear{2021}
% \setcopyright{rightsretained}
% \acmConference[L@S '21]{Proceedings of the Eighth ACM Conference on Learning @ Scale}{June 22--25, 2021}{Potsdam, Germany}
% \acmBooktitle{Proceedings of the Eighth ACM Conference on Learning @ Scale (L@S '21), June 22--25, 2021, Potsdam, Germany}\acmDOI{10.1145/3430895.3460139}
% \acmISBN{978-1-4503-8215-1/21/06}

\maketitle

\begin{abstract}
Early identification of college dropouts can provide tremendous value for improving student success and institutional effectiveness, and predictive analytics are increasingly used for this purpose. However, ethical concerns have emerged about whether including protected attributes in these prediction models discriminates against underrepresented student groups and exacerbates existing inequities. We examine this issue in the context of a large U.S. research university with both residential and fully online degree-seeking students. Based on comprehensive institutional records for the entire student population across multiple years (N = 93,457), we build machine learning models to predict student dropout after one academic year of study and compare the overall performance and fairness of model predictions with or without four protected attributes (gender, URM, first-generation student, and high financial need). We find that including protected attributes does not impact the overall prediction performance and it only marginally improves the algorithmic fairness of predictions. These findings suggest that including protected attributes is preferable. We offer guidance on how to evaluate the impact of including protected attributes in a local context, where institutional stakeholders seek to leverage predictive analytics to support student success.
\end{abstract}

% ACM Classfication



% Author Keywords

% Print the classficiation codes
% Please use the 2012 Classifiers and see this link to embed them in the text: \url{https://dl.acm.org/ccs/ccs_flat.cfm}

\section{Introduction}
% 1. Universities and online programs use predictive models to allocate resources and help at-risk students before they drop out
% 2. These models are built with a variety of data sources, which can include socio-demographic variables (protected attributes)
% 3. Scholars have raised questions about the inclusion of protected attributes in these models
% 4. We investigate the consequences of including protected attributes on model accuracy and fairness in a real-world scenario 
% RQ: How does the accuracy and fairness of college student at-risk predictive models change with the inclusion of protected attributes?

With the rapid development of learning analytics in higher education, data-driven instructional and learning support systems are increasingly adopted in classroom settings, and institution-level analytics systems are used to optimize resource allocation and support student success on a large scale. A common objective of these systems is the early identification of at-risk students, especially those likely to drop out of college. This type of prediction has significant policy implications because reducing college attrition has been a central task for institutional stakeholders ever since higher education was made accessible to the general public~\cite{pantages1978}. As of 2018, fewer than two-thirds of college students in the United States graduated within six years, and this share is even smaller at the least selective institutions which serve disproportionately more students from disadvantaged backgrounds~\cite{2020condition}. At the same time, the supply of academic, student affairs, and administrative personnel is insufficient to provide just-in-time support to students in need~\cite{2020condition}. It is within these resource-strained contexts that predicting dropouts based on increasingly digitized institutional data has the potential to augment the capacity of professionals who work to support student retention and success. Starting with the Course Signals project at Purdue University, an increasing number of early warning systems have explored this possibility at the institutional level~\cite{Arnold2012,Jayaprakash2014,Ekowo2016,Dawson2017}.

Accurately forecasting which students are likely to drop out is essentially profiling students based on a multitude of student attributes. These attributes often include socio-demographic information that is routinely studied in higher education research. Although the analysis of historical socio-demographic gaps in retention and graduation rates is well established in higher education research~\cite{2019racialtrends}, it becomes controversial to use these same characteristics when making predictions about the future. For example, is it fair to label a black first-year student as \textit{at risk} based on the higher dropout rate among black students in previous cohorts? The answer may be equivocal~\cite{sbs}. On the one hand, the observed historical gaps capture systematic inequalities in the educational environment of different student groups, which may well apply to future students from the same groups and therefore contribute to similar gaps. In this sense, explicitly using socio-demographic data can result in more accurate predictions and improve the efficiency of downstream interventions and actions based on those algorithmic decisions~\cite{Paquette2020}. On the other hand, from an ethics and equity perspective, the inclusion of socio-demographic variables may lead to discriminatory results if predictive models systematically assign differential predicted values across student groups based on the records of their historical counterparts. When these results are used for decision-making, stigmas and stereotypes could carry over to future students and reproduce existing inequalities~\cite{kizilcec2020,Barocas2019}.
%In order to close achievement gaps, it is thus desirable to keep predictive algorithms blind to demographics, in hopes that students are granted equal opportunities regardless of their background and identity.

In this paper, we investigate the issue of using protected attributes in college dropout prediction in real-world contexts. Protected attributes are traits or characteristics based on which discrimination is prescribed as illegal, such as gender, race, age, religion, and genetic information. We examine students in a residential college setting as well as students in fully online degree programs, which have been increasingly represented in formal higher education. In Fall 2018, 16.6\% of postsecondary students in the United States were enrolled in exclusively online programs, up from 12.8\% in Fall 2012~\cite{Seaman2018,2019digest}. The absence of a residential experience exposes students to additional challenges to accountability and engagement, and also makes it harder for faculty and staff members to identify problems with students' well-being and provide timely support. The COVID-19 pandemic has forced most colleges to move instruction online, which will likely increase the importance of online learning in the future of higher education~\cite{chronicle}. Predictive analytics are therefore just as useful for online higher education as they are for residential settings for supporting student achievement and on-time graduation. Our findings in both residential and online settings offer practical implications to a broad range of stakeholders in higher education.

By systematically comparing predictive models with and without protected attributes in two higher education contexts, we aim to answer the following two research questions:
\begin{enumerate}
    \item How does the inclusion of protected attributes affect the overall performance of college dropout prediction? 
    \item How does the inclusion of protected attributes affect the fairness of college dropout prediction?
    % \item How can we evaluate the trade-off between accuracy and fairness in college dropout prediction with respect to protected attributes?
\end{enumerate}

This research contributes to the literature on predictive modeling and algorithmic fairness in (higher) education on several dimensions. First, we present one of the largest and most comprehensive evaluation studies of college dropout prediction based on student data over multiple years from a large public research university. % using a realistic feature set drawn from student information systems.
This offers robust insights to researchers and institutional stakeholders into how these models work and where they might go wrong. Second, we apply the prediction models with the same features to both residential and online degree settings, which advances our understanding of generalizability across contexts, such as in which environment it is easier to predict dropout and to what degree key predictors differ.
%the relative difficulty of predicting dropout in these contexts and which features matter where. 
Third, we contribute some of the first empirical evidence on how the inclusion of protected attributes affects the fairness of dropout prediction, which can inform equitable higher education policy around the use of predictive modeling. %in terms of three measures of fairness and four student characteristics.
%Our findings therefore provide robust evidence to inform higher education policy on predictive modeling.

\section{Related Work}
\label{sec:lit}

\subsection{College Dropout Prediction}
\label{sec:lit:droppred}

Decades of research have charted the ecosystem of higher education as a complex journey with "a wide path with twists, turns, detours, roundabouts, and occasional dead ends that many students encounter" and jointly shape their academic and career outcomes~\cite{Kuh2007}. Among the variety of factors that influence students' journey, background characteristics such as demographics, family background, and prior academic history are strong signals of academic, social, and economic resources available to a student before adulthood, which are substantially correlated with college success~\cite{coleman1988social}. For example, ethnic minorities, students from low-income families, and first-generation college students have consistently suffered higher dropout rates than their counterparts~\cite{2019racialtrends,2018firstgen}, and students who belong to more than one of these groups are even more likely to drop out of college. In addition to these largely immutable attributes at college entry, students' experiences in college such as engagement and performance in academic activities are major factors for success. In particular, early course grades are among the best predictors of persistence and graduation, even after controlling for background characteristics~\cite{Kuh2007}.

With the advent of the "datafication" of higher education~\cite{Selwyn2020}, there has been an increasing thrust of research to translate the empirical understanding of dropout risk factors into predictive models of student dropout (or success) using large-scale administrative data \cite{Aulck2019,Dekker2009,Jayaprakash2014,DelBonifro2020,Beaulac2019,Berens2019,Hutt2019}. These applications are usually intended to facilitate targeted student support and intervention programs, and the extensive research literature on college success has facilitated feature engineering grounded in theory. %Given the importance of academic activities and the universal availability of students' academic records at institutions, recent learning analytics work has built predictive pipelines on large-scale administrative data to identify at-risk students early~\cite{Aulck2019,Arnold2012,Jayaprakash2014,Beaulac2019,Berens2019,Hutt2019}. This offered some balance between model interpretability and prediction accuracy: 
For example, Aulck and colleagues~\cite{Aulck2019} used seven groups of freshman features extracted from registrar data to predict outcomes for the entire student population at a large public university in the US. The model achieved an accuracy of 83.2\% for graduation prediction and 95.3\% for retention. In a more application-oriented study as part of the Open Academic Analytics Initiative (OAAI), Jayaprakash and colleagues~\cite{Jayaprakash2014} developed an early alert system that incorporated administrative and learning management system data to predict at-risk students (those who are not in good standing) at a small private college, and then tested the system at four other less-selective colleges.

While the recent decade has seen a steady growth in prediction-focused studies on college dropout, a large proportion of them are focused on individual courses or a small sample of degree programs~\cite{Hellas2018}. Most of them investigate dropouts at brick-and-mortar institutions. Our study pushes these research boundaries by examining dropout prediction for multiple cohorts of students across residential and exclusively online degree programs offered by a large public university. The breath of our sample is rare in the dropout prediction literature and promises to offer more generalizable insights about the utility and feasibility of predictive models.
%Prior research suggests that students enrolled in online programs are more likely to be non-traditional students and come from historically underrepresented groups \cite{2019digest}. The challenges associated with the online format expose these students to additional risks for dropping out of the program \cite{Xu2020}. 

\subsection{Algorithmic Fairness in Education}
\label{sec:lit:algfair}

% In recent years, machine learning and algorithmic systems have been gradually applied to high-stakes contexts, such as in criminal justice and healthcare, to facilitate decision making or even take control over human decision makers. This trend has called into question the issues of algorithmic bias and fairness, given how inaccuracies in the models might translate into severe consequences in the real world \cite{Barocas2019}. Following legal terms, algorithmic fairness is generally evaluated with respect to individual protected attributes. The specific criteria of fairness, however, vary and largely depend on the purpose of the specific application(s) \cite{verma2018}.

A central goal of educational research and practice has been to close opportunity and achievement gaps between different groups of students. More recently, algorithmic fairness has become a topic of interest as an increasing number of students are exposed to intelligent educational technologies~\cite{kizilcec2020}. Inaccuracies in models might translate into severe consequences for individual students, such as failing to allocate remedial resources to struggling learners. It is more concerning if such inaccuracies disproportionately fall upon students from disadvantaged backgrounds and worsen existing inequalities. In this context, the fairness of algorithmic systems is generally evaluated with respect to protected attributes following legal terms. The specific criteria of fairness, however, vary and largely depend on the specific application(s) \cite{verma2018}.

In the past few years, a handful of papers have brought the fairness framework to real-world learning analytics research. Most of these studies audit whether supervised learning models trained on the entire student population generate systematically biased predictions of individual outcomes such as correct answers, test scores, course grades, and graduation~\cite{Yu2020,Kung2020,Gardner2019,Hutt2019,Doroudi2019,loukina2019}. For example, Yu and colleagues~\cite{Yu2020} found that models using college-entry characteristics to predict course grades and GPA tend to predict lower values for underrepresented student groups than their counterparts. Other studies have examined biases encoded in unsupervised representations of student writing~\cite{arthurs2020}, or go further to refine algorithms for at-risk student identification under fairness constraints~\cite{Hu2020}. Overall, this area of research is nascent and in need of systematic frameworks specific to educational contexts to map an agenda for future research.

When it comes to strategies to improve algorithmic fairness, a contentious point is whether protected attributes should be included as predictors (features) in prediction models. Most training data from the real world are the result of historical prejudices against certain protected groups, so directly using group indicators to predict outcomes risks imposing unfair stereotypes and reproduce existing inequalities~\cite{Barocas2019}. In educational settings, it may be considered unethical to label students from certain groups as "at risk" from day one, when in fact, these students have demonstrated an exceptional ability to overcome historical obstacles and might therefore be more likely to succeed~\cite{sbs}. This concern motivated the research effort to ``blind'' prediction models by simply removing protected attributes (i.e. fairness through unawareness) or more complicated statistical techniques to disentangle signals of protected attributes from other features due to their inherent correlation~\cite{calmon2017}. In contrast, recent work has advocated for explicitly using protected attributes in predictive models (i.e. fairness through awareness)~\cite{dwork2012}. In particular, Kleinberg and colleagues~\cite{Kleinberg2018} showed in a synthetic example of college admission that the inclusion of race as a predictor of college success improves the fairness of admission decisions without sacrificing efficiency. Given the well-documented relationship between student background and their educational outcomes, a recent review also suggests that predictive models in education should include demographic variables to ensure that algorithms are value-aligned, i.e., all students have their needs met~\cite{Paquette2020}.

To our knowledge, however, there is only limited empirical evidence to support either side of this debate. Our study therefore presents an in-depth examination of the consequences of including or excluding protected attributes on algorithmic fairness of a realistic, large-scale dropout prediction model.

\section{Methodology}

\begin{table*}[t]
  \centering
  \caption{Features used for dropout prediction.}~\label{tab:features}
  \begin{tabular}{p{0.2\linewidth} p{0.7\linewidth}}
    \toprule
    Category & Features\\
    \midrule
    Protected attributes & Gender (binary), first-generation college student (binary), underrepresented minority (URM; binary; defined as not Asian or White), high financial need (binary; FASFA-based expected family contribution under \$5,500)\\
    Incoming attributes & Age, high school GPA, math and verbal SAT/ACT scores, transfer student (binary), transferred credits, transfer GPA\\
    Program information & Part-time student (binary), major, minor, STEM major (binary)\\
    Course performance & Total courses enrolled, total units enrolled, percentage of courses that are required, credits received from different types of courses (lecture, seminar, etc.), levels of courses (100, 200, etc.), term GPA, mean and variance in course grades within each session during the term, percentage distribution of letter grades\\
  \bottomrule
  \end{tabular}
\end{table*}

\subsection{Dataset}

We analyze de-identified institutional records from one of the largest public universities in the United States. This broad-access research university serves nearly 150,000 students with an 86\% acceptance rate and 67\% graduation rate. Its student population is representative of the state in which it is located, which makes it a Hispanic-serving institution (HSI). The university has offered many of the same undergraduate degree programs fully online to over 40,000 students. The dataset we use in this study focuses on undergraduate students and contains student-level characteristics and student-course-level records for their first term of enrollment at the university, including transfer students (except for those who transfer into their senior year). For our prediction task, we only keep students whose first term was in the Fall along with their course-taking records in their first term, including terms between 2012-18 (residential) and 2014-18 (online).

This sample comprises a total of 564,104 residential course-taking records for 93,457 unique students and 2,877 unique courses, and 81,858 online course-taking records for 24,198 unique students and 874 unique courses. The course-taking records include both a student's letter grade and course-level metadata (subject, course number, units, required for major, etc.). Student-level information includes socio-demographic information (age, gender, race/ethnicity, first-generation status, etc.), prior academic achievement (high school GPA, standardized test scores), enrollment information (transfer student status, part-time status, academic major and minor, etc.). These data are representative of what most higher education institutions routinely manage in their student information systems (SIS)~\cite{Aulck2019}.

\subsection{Prediction Target and Feature Engineering}

The primary goal of a dropout prediction model is to alert relevant stakeholders to currently enrolled students who are at risk of dropping out of a degree program so that they can reach out and offer support at an early stage. While the general framework of dropout prediction is well established, the exact definition of dropout, or attrition, varies based on the specific context~\cite{pantages1978}. In our context, we define dropout as not returning to school a year from the first time of enrollment. We only analyze students who first enrolled in Fall, so dropout means not returning in the following Fall. This final operationalization aligns well with retention, one of the two standard metrics of post-secondary student success in national reports of the United States \cite{2019digest,2020condition}.\footnote{The other standard metric is graduation within 100\% or 150\% of the normative time (i.e. 4 or 6 years for four-year institutions). We do not examine this metric because the span of our dataset is only six years and we do not observe graduation outcomes for all student cohorts.}

We use students' background characteristics and academic records in the first enrolled term (Fall) to predict dropout, because it would be beneficial to identify risks as early as possible and institutional records are usually updated and available at the end of each term. Informed by existing research in higher education and learning analytics (see Related Work), we construct 58 features from the dataset for both residential and online students. Table~\ref{tab:features} summarizes these feature by four categories. We include four protected attributes, which are the most commonly used dimensions along which to examine educational inequalities and set equity goals in policy contexts~\cite{2018firstgen,chen2019profile,2020condition}.

Table~\ref{tab:data_profile} depicts the student profile in our analysis. The statistics reaffirm that, regardless of format, the institution serves a large proportion of students from historically disadvantaged groups. There are also major differences across formats. In line with the national statistics of exclusively online programs~\cite{2019digest}, the online sample has a higher concentration of transfer and non-traditional (older, part-time) students, and also higher dropout rates compared to residential students. These characteristics validate that the current analysis is performed on student populations who are most in need of institutional support and allow us to scrutinize the generalizability of our findings across two distinct contexts of higher education.

\begin{table}[h]
\centering
\caption{Comparison of online and residential student populations.}~\label{tab:data_profile}
\begin{tabular}{lcc}
\toprule
 & Online & Residential \\ 
 \midrule
N & 24,198 & 93,457 \\
Dropout & 40.7\% & 16.9\% \\
Female & 60.9\% & 47.9\% \\
First-gen & 42.4\% & 33.6\% \\
URM & 33.1\% & 34.6\% \\
High need & 61.9\% & 51.3\% \\
Transfer & 85.2\% & 31.8\% \\
Part-time & 77.2\% & 12.9\% \\
% Cumulative GPA & 2.51 & 3.05 \\
Average age & 27.1 & 19.7 \\
\bottomrule
\end{tabular}
\end{table}

\subsection{Dropout Prediction}

To investigate the consequences of using protected attributes in dropout prediction models, we generate two feature sets: the AWARE set includes all features shown in Table~\ref{tab:features}, while the BLIND set excludes the four protected attributes from the AWARE set. For convenience, we will refer to a specific model by the feature set it uses in the remainder of this paper. %Note that we opt not to compare against a model with pre-processed data. While some scholars have suggested further reducing signals of protected attributes that are embedded in the remaining features using pre-processing techniques (e.g., removing gender and racial bias in standardized test scores) \cite{calmon2017}, these approaches remain an active area of research and have not yet made their way into common applications.
Given our binary target variable, the dropout prediction task is formalized as a binary classification problem. As we focus on identifying the effect of including protected attributes, we experiment with two commonly used algorithms -- logistic regression (LR) and gradient boosted trees (GBT). We choose LR because it is a linear additive and highly interpretable classifier that can achieve reasonable prediction performance with well-chosen features. The choice of GBT, on the other hand, is for its ability to accommodate a large number of features, efficiently handle missing values, and automatically capture non-linear interactions between features.
%of the comparatively large number of features, the presence of missing values, and the explicit interest in testing potential interactions between protected attributes and other features in the AWARE model. 

We predict dropping out separately for online and residential students. For each format, we split the data into a training set and a test set based on student cohort: the last observed cohort (6,939 online and 14,275 residential students entering in Fall 2018) constitutes the test set and the remaining cohorts make up the training set (17,259 online and 79,182 residential students). There are two reasons for doing the train-test split by student cohorts. Practically, this split aligns with the real-world application where stakeholders rely on historical data to make predictions for current students~\cite{Jayaprakash2014}. Technically, this approach alleviates the issue of data contamination between the training and test set~\cite{farrow2019analysing}, as the features we use, especially the first-semester records, might be highly correlated within the same cohort but much less so across cohorts. 

There are a few additional technical details about model training. First, we tune hyperparameters of the two algorithms by performing grid search over a specified search space and evaluating the hyperparameters using 5-fold cross-validation. Second, we add indicator variables for missing values in course grades, standardized test scores, and academic majors and minors. Third, we apply robust scaling to training features to regulate the influence of outliers. Fourth, because the class imbalance in both datasets can bias the model learning towards the majority class (i.e. non-dropout), we adjust the sample weights to be inversely proportional to class frequencies during the training stage.

The trained classifiers are then applied to the test set to evaluate the performance. The immediate output of each classifier is a predicted probability of dropping out for each student. To make a final binary prediction of dropout, we use dropout rates in the training data to determine the decision thresholds for the test set, such that the proportion of predicted dropouts in the test set matches the proportion of observed dropouts in the training set~\cite{Berens2019}. Compared to the default of 0.5, this choice of threshold is more reasonable when we rely on the observed history to predict the unknown future in practice.

\subsection{Performance Evaluation}

We evaluate prediction performance based on three metrics: accuracy, recall, and true negative rate (TNR). In the context of dropout prediction, recall is the proportion of actual dropouts who are correctly identified, whereas TNR quantifies how likely a student who persists into the second year of college is predicted to persist.
% The following equations mathematically present the three metrics:
% \begin{equation}
%     Acc = P\left(\hat{Y}=d\middle|Y=d\right)
% \end{equation}
% \begin{equation}
%     Recall = P\left(\hat{Y}=1\middle|Y=1\right)
% \end{equation}
% \begin{equation}
%     TNR = P\left(\hat{Y}=0\middle|Y=0\right)
% \end{equation}
% These metrics can inform different aspects of policy considerations when adopting a dropout prediction application.
To examine the effects of including protected attributes on overall performance, we compute these metrics separately for each model and test whether each metric significantly changes from BLIND to AWARE models, using two proportion $z$-tests.

We operationalize fairness as the independence between prediction performance, measured by the three metrics above, and protected group membership. This definition of fairness with respect to the three metrics corresponds to the established notions of overall accuracy equality, equal opportunity, and predictive equality, respectively~\cite{kizilcec2020}. Specifically, to quantify the fairness of a given model with regard to a binary protected attribute, such as URM, we compute the differences in each of the three metrics between the two associated protected groups, URM and non-URM students. We then compare how much these differences change between BLIND and AWARE models in order to quantify the effect of including protected attributes as predictors on fairness.

\section{Results}

\subsection{Overall Prediction Performance}

\begin{table}[t]
\centering 
  \caption{Overall prediction performance of AWARE and BLIND models trained with gradient boosted trees (GBT) and logistic regression (LR).}
\small
  \begin{tabular}{lcccccc}
    \toprule
     & \multicolumn{3}{c}{GBT} & \multicolumn{3}{c}{LR}\\
    \cmidrule(lr){2-4} \cmidrule(lr){5-7}
        Metric & AWARE & BLIND & $\Delta$ & AWARE & BLIND & $\Delta$ \\
    \midrule
        \multicolumn{7}{l}{\textit{Online (Non-dropout: 59.3\%)}}\\
    Accuracy & 75.8 & 75.6 & 0.2 & 75.2 & 75.4 & -0.2\\
    Recall & 67.3 & 67.1 & 0.2 & 66.7 & 66.8 & -0.1\\
    TNR & 82.4 & 82.3 & 0.1 & 81.9 & 82.0 & -0.1\\ 
    \midrule
        \multicolumn{7}{l}{\textit{Residential (Non-dropout: 83.1\%)}}\\
    Accuracy & 83.9 & 83.9 & 0.0 & 83.6 & 83.6 & 0.0\\
    Recall & 54.1 & 54.1 & 0.0 & 53.2 & 53.3 & -0.1\\
    TNR & 89.1 & 89.1 & 0.0 & 88.9 & 88.9 & 0.0\\
        \bottomrule
    \multicolumn{7}{l}{Note: none of the $\Delta$ values is statistically significant with $p<0.1$.}
  \end{tabular}
\label{tab:overall_perf}
\end{table}

\begin{figure*}[h]
    \centering
    \begin{minipage}{0.5\textwidth}
        \centering
        \includegraphics[width=\textwidth]{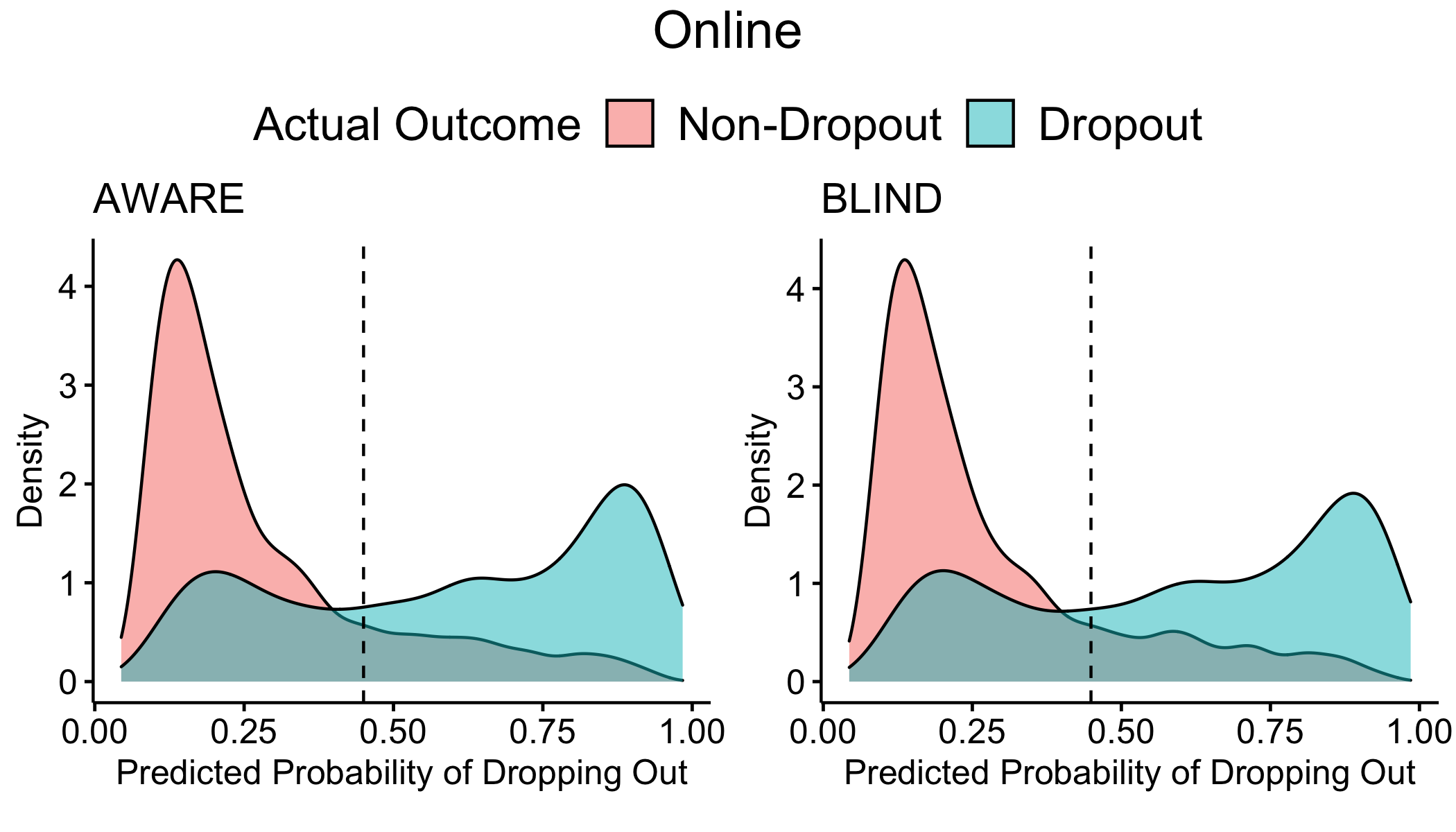}
    \end{minipage}\hfill
    \begin{minipage}{0.5\textwidth}
        \centering
        \includegraphics[width=\textwidth]{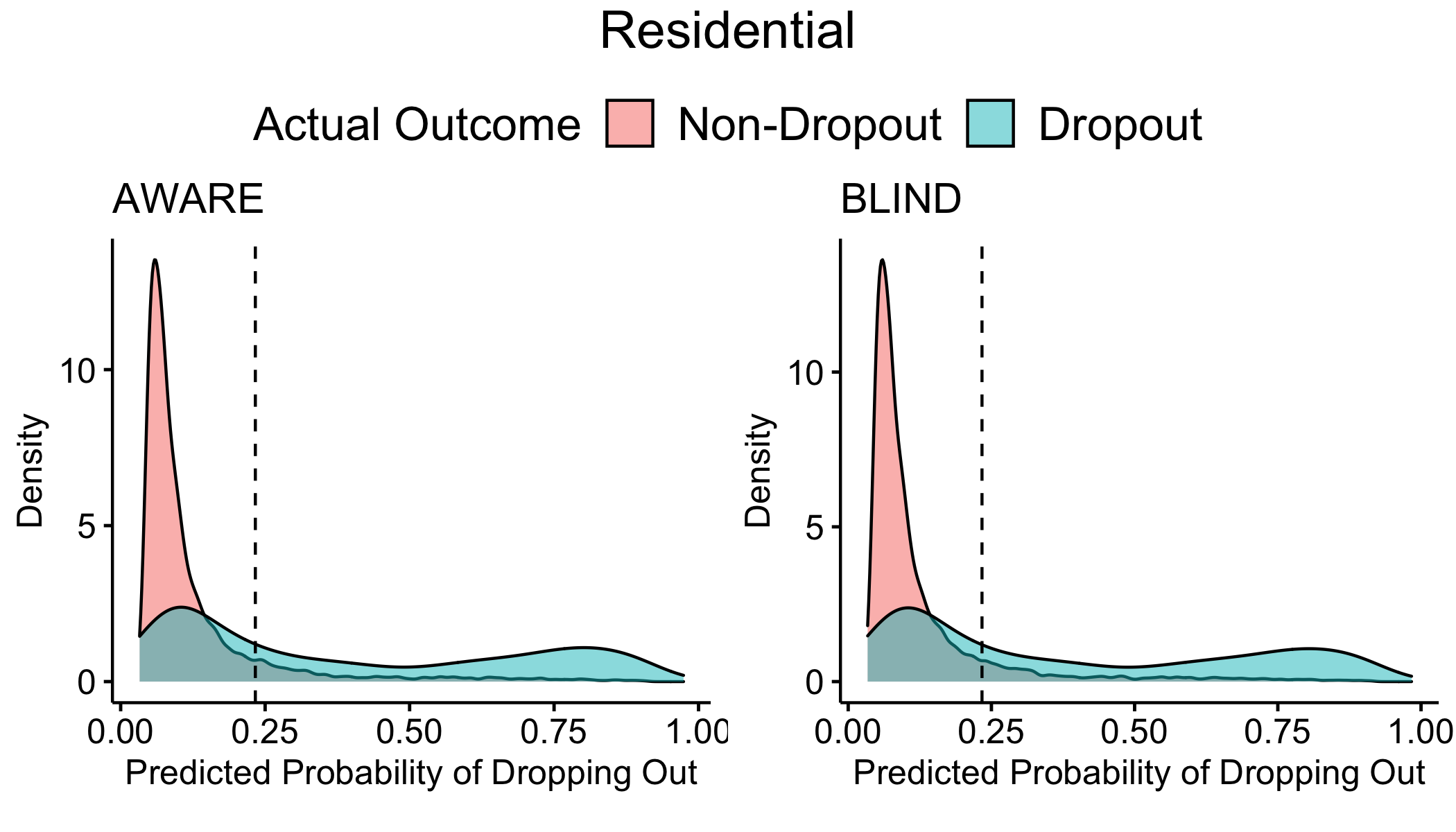}
    \end{minipage}\hfill
  \caption{Distribution of predicted dropout probability}
  \label{fig:overall_prob_dist}
\end{figure*}

We first illustrate the effects of including protected attributes on overall prediction performance. Table~\ref{tab:overall_perf} reports the overall performance of AWARE and BLIND models, trained with GBT and LR algorithms, on the test dataset. The last column under each algorithm reports the percentage point differences in performance between the two models (from BLIND to AWARE). The main finding is that including or excluding protected attributes does affect the performance of the dropout prediction in either context. None of the performance metrics (accuracy, recall, TNR) differs significantly between the BLIND and AWARE models. Additionally, while the more sophisticated GBT algorithm performs better than the simple LR on all metrics, the advantage is comparatively small (less than one percentage point on all metrics). Because of this, we restrict the following analysis to GBT-based models.

Compared to a na\"ive baseline which simply predicts every student to be the majority class (non-dropout) and achieves an accuracy equal to that majority's share, the predictive models can accurately predict online dropouts with a decent margin. However, the accuracy margin for predicting residential dropouts is fairly small. The other two metrics, which describe the accuracy among dropouts and non-dropouts respectively, achieve a higher value when the corresponding group has a larger share and vice versa. Specifically, the models are able to identify 67.3\% of online dropouts and 54.1\% of residential dropouts. This latter value is somewhat lower but still comparable to the recall performance in recent prior work on dropout prediction in residential programs~\cite{Berens2019,DelBonifro2020}.
% \begin{table}
% \centering 
%   \caption{Overall prediction performance of the AWARE And BLIND model for online and residential students.}
%   \begin{tabular}{lcccccc}
%     \toprule
%     \multicolumn{1}{l}{Model} & \multicolumn{3}{c}{GBT} & \multicolumn{3}{c}{LR}\\
%     Metric & AWARE & BLIND & $\Delta$ & AWARE & BLIND & $\Delta$
%     \midrule
%     \multicolumn{7}{l}{\textit{Online Students}}\\
%     Accuracy & 75.8 & 75.6 & 0.2 & 75.2&75.4 & -0.2\\
%     Recall & 67.3 & 67.1 & 0.2 & 66.7& 66.8&-0.1\\
%     TNR & 82.4 & 82.3 & 0.1 & 81.9& 82.0& -0.1\\
%     \midrule
%     \multicolumn{7}{l}{\textit{Residential Students}}\\
%     Accuracy & 83.9 & 83.9 & 0.0 & 83.6& 83.6& 0\\
%     Recall & 54.1 & 54.1 & 0.0 & 53.2& 53.3& -0.1\\
%     TNR & 89.1 & 89.1 & 0.0 & 88.9& 88.9& 0.0\\
%     \bottomrule
%      %\multicolumn{4}{l}{Sig. levels: ***0.01; **0.05; *0.1}
%   \end{tabular}
% \label{tab:overall_perf}
% \end{table}

To take a closer look at the model predictions, beyond the three aggregate performance metrics, we examine whether including protected attributes alters the distribution of predicted dropout probabilities. As shown in Figure~\ref{fig:overall_prob_dist}, the distributions are highly similar across the models which further validates the limited marginal impact of protected attributes. An additional insight from these plots is that dropouts might be much more heterogeneous than non-dropouts in terms of the features in Table~\ref{tab:features}, as their predicted probabilities are highly spread out, especially in residential settings where the majority of dropouts are assigned a small dropout probability. This pattern is consistent with the lower recall performance shown in Table~\ref{tab:overall_perf}.

% \begin{figure*}[h]
%   \centering
%   \includegraphics[width=0.8\linewidth]{plots/overall_probability_density_onl.png}\hfill
%   \includegraphics[width=0.8\linewidth]{plots/overall_probability_density_f2f.png}
%   \caption{Distribution of predicted dropout probability in AWARE (left) and BLIND (right) models, for online (top) and residential (bottom) students.}
%   \label{fig:overall_prob_dist}
%   \Description{}
% \end{figure*}

% Second, we take the explanatory perspective of feature importance and evaluate if protected attributes assume decent importance when included in the model. Figure~\ref{fig:varimp} shows ten most important features, measured by their contribution to the mean decrease in Gini, in the AWARE and BLIND models. We find that, not only protected attributes fail to show up in this list for the AWARE model, but also the same ten features top the leaderboard in exactly the same order of importance. This pattern lends further credit to the inability of protected attributes to help more accurately predict dropouts.

% \begin{figure*}[h]
%   \centering
%   \includegraphics[width=0.9\linewidth]{plots/fig_imp.png}
%   \caption{Ten most important features in AWARE and BLIND models}
%   \label{fig:varimp}
%   \Description{}
% \end{figure*}

This finding appears to conflict with prior research that demonstrates the critical role of demographic and background characteristics for student success in higher education~\cite{Kuh2007}. In an effort to better understand our result, we explore two mutually compatible hypotheses inspired by the algorithmic fairness literature. One hypothesis is that dropping out, the prediction target, is not sufficiently correlated with protected attributes, and thus adding the latter to a dropout prediction model would not improve performance much. To test this, we fit separately for each enrollment format in the test data a logistic regression model that predicts dropout using all the possible interaction terms between the four protected attributes. We find that, even though a few coefficients are statistically significant, the adjusted McFadden’s $R^2$ is as small as 0.006 for either format, lending support to our hypothesis.

The second hypothesis is that protected attributes are already implicitly encoded in the BLIND feature set, and adding them directly does not add much predictive power. We test this by fitting four logistic regressions for each format which use the BLIND feature set to predict each of the four protected attributes. Based on the adjusted McFadden’s $R^2$, we find that only gender can plausibly be considered encoded in the other features (0.159 for online and 0.187 for residential). This lends partial support to our second hypothesis.

% c) orthogonalize covars
% -- compare results after vs. before ortho -> assumption 1)
% -- after ortho, compare results w/ and w/o PAs -> assumption 2)

\subsection{Fairness of Prediction}

\begin{figure*}[h]
  \centering
  \includegraphics[width=0.5\linewidth]{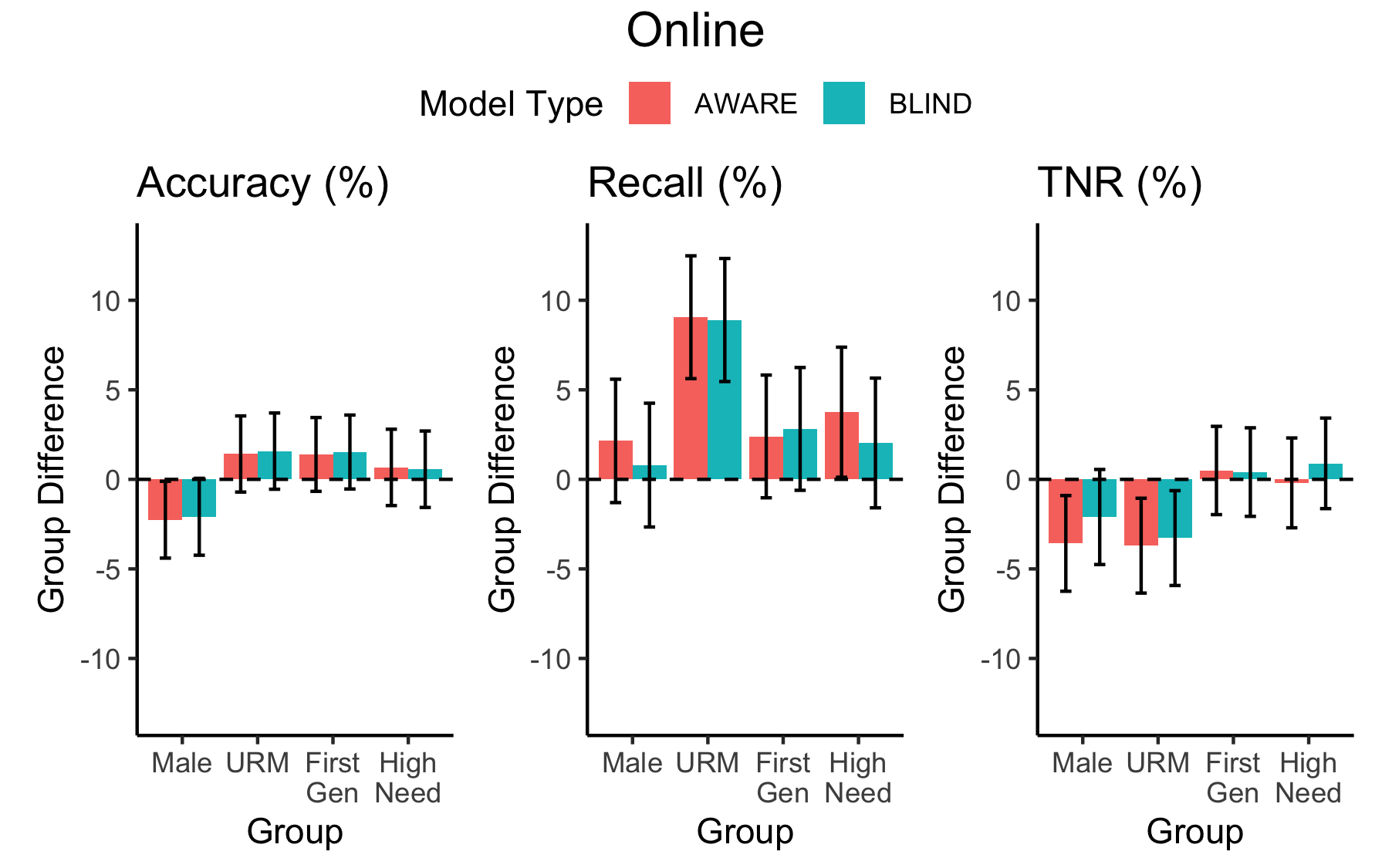}\hfill
  \includegraphics[width=0.5\linewidth]{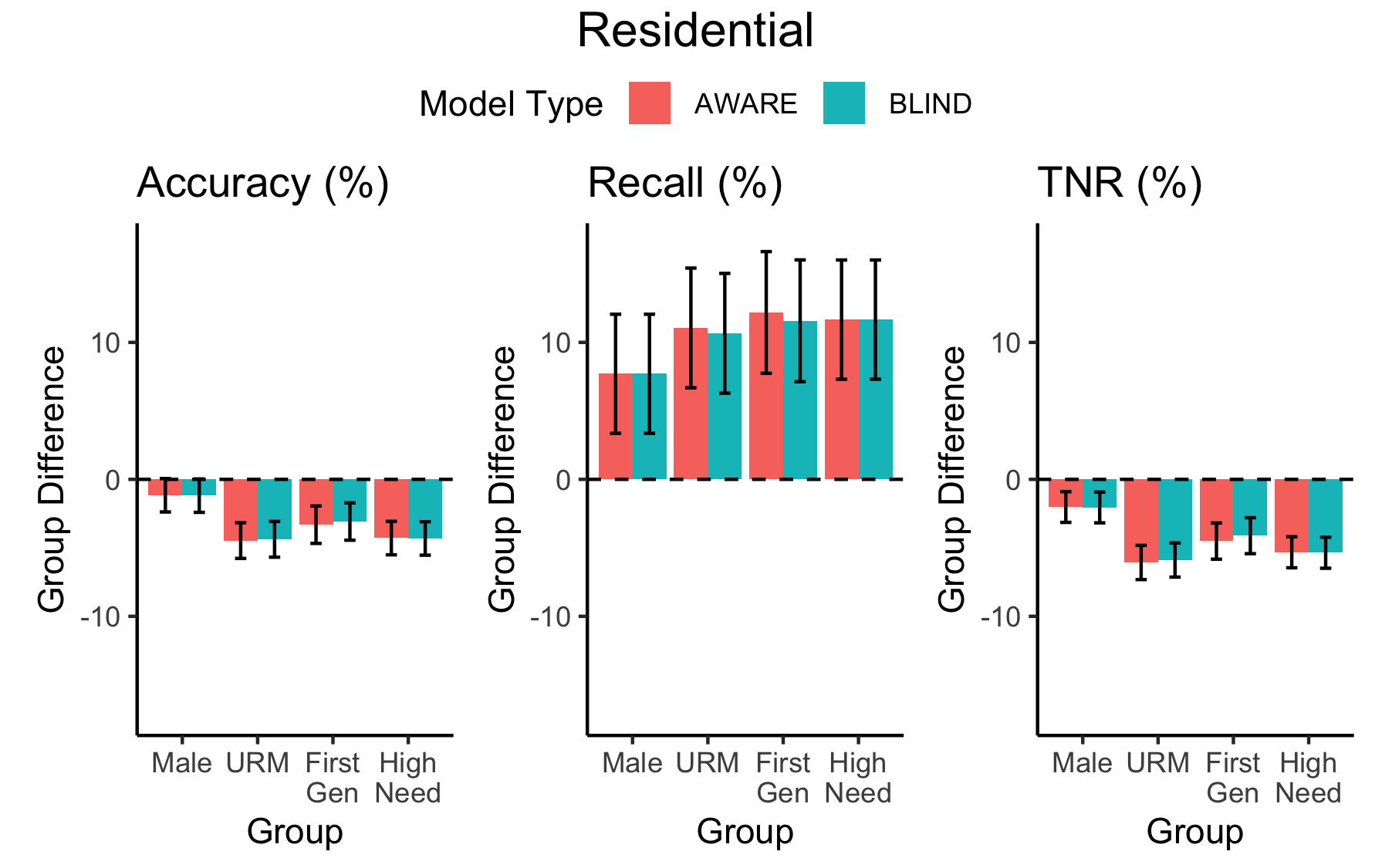}\hfill
  \caption{Fairness of AWARE and BLIND models in terms of accuracy (left), recall (middle), and TNR (right). Positive group differences (y-axis) indicate higher values for the listed groups compared to their corresponding reference groups. Group differences closer to zero reflect higher levels of fairness. Error bars indicate 95\% confidence intervals.}
  \label{fig:fairness}
\end{figure*}

% \begin{table}
% \centering
% \caption{Dropout Rates Among Different Protected Groups in the Test Sample for Online (Top) and Residential (Bottom) Students Students}
% \begin{tabular}{@{}ccc@{}}
% \toprule
% Proportion (\%) & Dropout & Non-Dropout \\ \midrule
%     \multicolumn{4}{l}{\textit{Online Students}}\\
% Male & 41.0 & 32.8 \\
% URM & 37.7 & 33.8 \\
% First Gen & 42.6 & 43.0 \\
% High Need & 66.4 & 61.3 \\
% \midrule
%     \multicolumn{4}{l}{\textit{Residential Students}}\\
% Male & 54.1 & 50.8 \\
% URM & 41.1 & 35.5 \\
% First Gen & 36.7 & 29.4 \\
% High Need & 56.9 & 47.8 \\
% \bottomrule
% \end{tabular}
% \label{tab:group_dropout_rate}
% \end{table}

We further examine how the inclusion of protected attributes might affect the fairness of dropout predictions. As mentioned in the previous section, for each of the four protected attributes, we first measure fairness by the group difference in a chosen performance metric. For example, a prediction model that achieves the same accuracy on male and female students is considered fair in terms of accuracy (0\% difference). Following this construction, Figure~\ref{fig:fairness} visualizes these fairness results of the AWARE and BLIND models for each of the four protected attributes in terms of the three metrics. Each bar in a subplot depicts the difference in that metric between the labeled group and their counterpart (e.g., male - female). The closer the bar is to zero, the fairer that model prediction is. Overall, the figure shows that both the AWARE and BLIND models are unfair for some protected attributes and some metrics, but fair for others. This lack of universal fairness is expected given the many dimensions of protected attributes, models, and metrics. However, for residential students, the model consistently exhibits unfairness across all protected attributes and metrics, especially in terms of recall. The inclusion or exclusion of protected attributes does not in general lead to different levels of fairness in terms of any metric in any enrollment format, as all adjacent error bars in the figure exhibit a high degree of overlap.

\begin{figure*}[h]
    \centering
    \begin{minipage}{0.5\textwidth}
        \centering
        \includegraphics[width=1\linewidth]{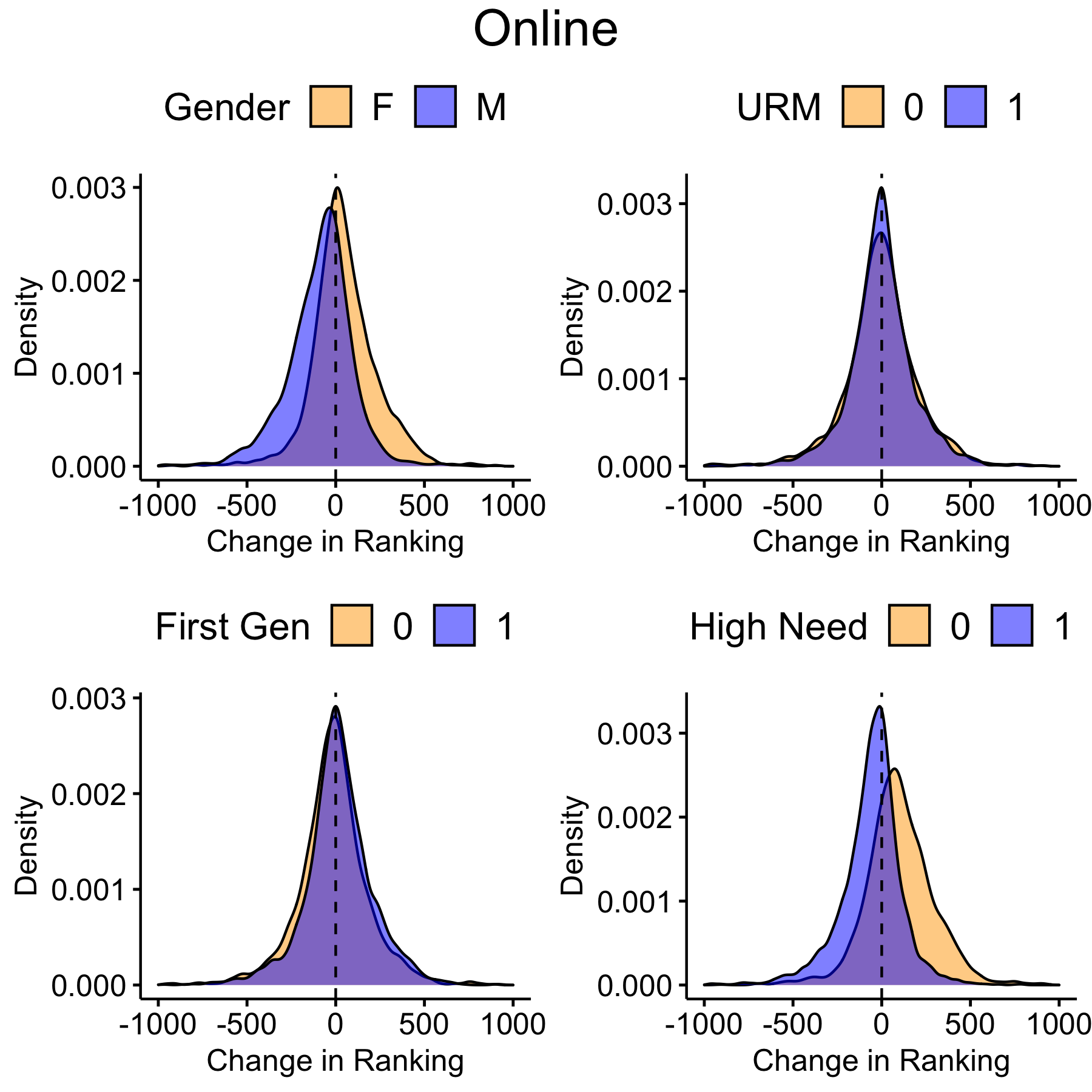}
    \end{minipage}\hfill
    \begin{minipage}{0.5\textwidth}
        \centering
        \includegraphics[width=1\linewidth]{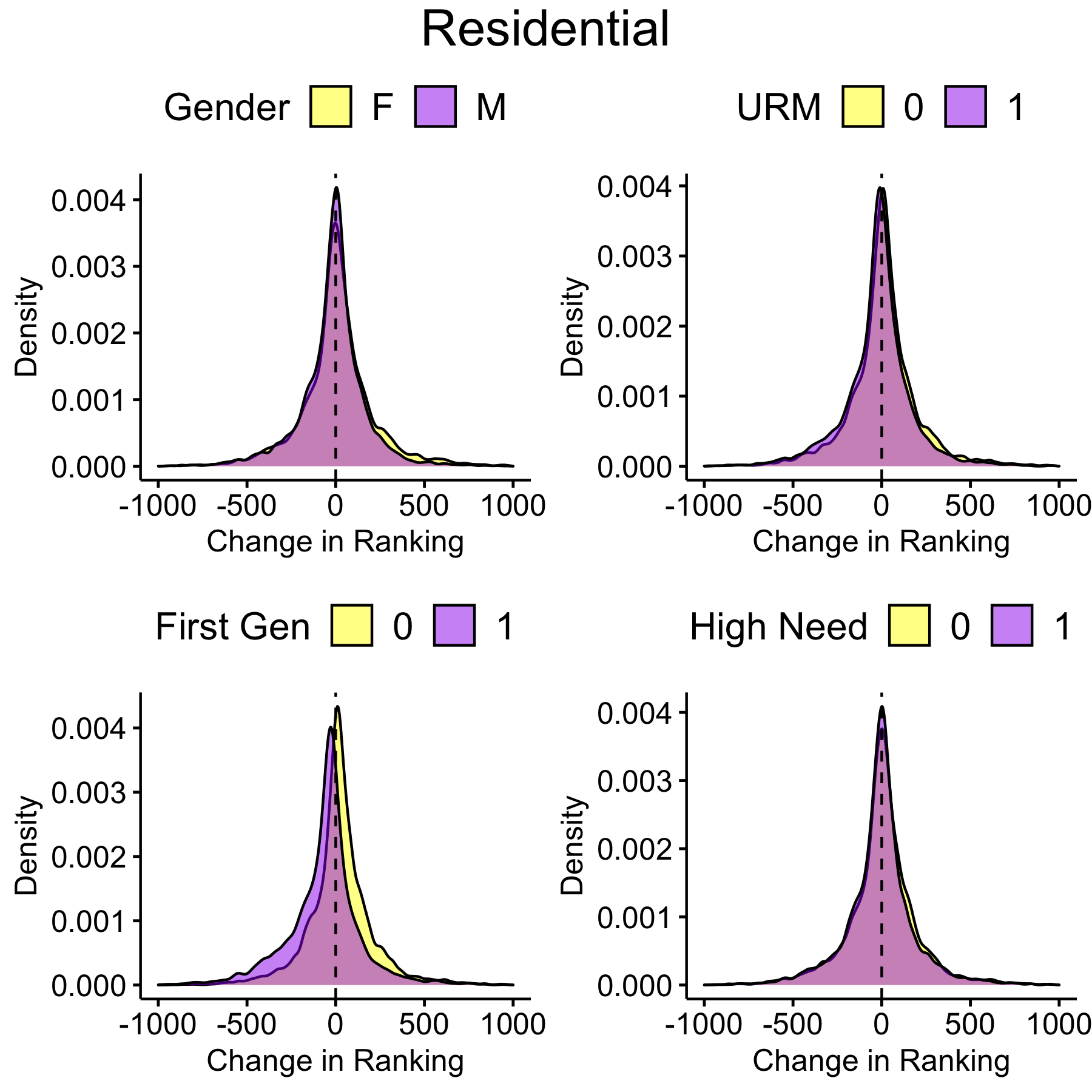}
    \end{minipage}\hfill
  \caption{Distribution of change in individual ranking of predicted dropout probability from AWARE to BLIND. One unit increase means going up by one place in the AWARE model compared to the BLIND model.}
  \label{fig:ranking}
\end{figure*}

While the aggregated group fairness metrics do not differ with vs. without protected attributes, we take a step further to explore how individual-level changes in model predictions can shed light on the overall change in fairness. We examine changes in the individual ranking of predicted dropout probability among all predicted students (test set) from BLIND to AWARE model. Figure~\ref{fig:ranking} plots the distribution of this ranking change for each protected group, where higher values represent moving up in the assigned risk leaderboard when protected attributes are included for prediction. 

We find that overall the ranking change is centered around zero, but there are observable group differences in certain cases. In the online setting, the AWARE model tends to move up females and students without a high financial need on the dropout risk leaderboard simply based on their identity. Similarly, continuing-generation college students are moved up more in residential settings compared to their first-generation counterparts. We argue that these group differences suggest improved fairness if the group going up more in the ranking spectrum has lower dropout rates in reality, and vice versa. To formally evaluate this reasoning, we conduct a series of $t$-tests between pairs of protected groups on their ranking change. We also compute Cohen's $d$ to gauge the standardized effect size. Comparing Table~\ref{tab:rankingchange} which describes these results and Table~\ref{tab:group_dropout_rate} which presents the actual dropout rates of each group, we find that moving from BLIND to AWARE causes students from advantaged (lower dropout rates) groups to be assigned relatively higher risk rankings compared to their disadvantaged (higher dropout rates) reference groups, and that this effect size is larger when the two paired groups have larger gaps in dropout rates. Thus, adding protected attributes to the model is working against existing inequities to a marginal extent instead of reinforcing them.

%The rationale is that if the actual lower-risk group goes up more in predicted risks than the actual higher-risk group, the model tends to close the gaps and therefore is more fair. For each model, we rank each student by their predicted dropout probability and then compute the ranking change for each student between BLIND and AWARE models. We test the difference in this ranking change between each pair of protected groups using independent t-tests and also calculate Cohen's d to evaluate the magnitude of that difference. Finally, we compare this group difference in ranking change with the same group difference in actual dropout rates. 

\begin{table}[t]
\centering
\caption{Dropout rates among different protected groups in the test set.}
\begin{tabular}{@{}lcc@{}}
\toprule
& \multicolumn{2}{c}{Dropout rate} \\
\cmidrule{2-3}
& Online & Residential \\ 
\midrule
Overall & 40.7 & 16.9 \\
\midrule
Male & 49.5 & 15.6 \\
Female & 40.7 & 14.0 \\
\midrule
URM & 46.6 & 16.8 \\
Non-URM & 42.4 & 13.7 \\
\midrule
First-gen & 43.7 & 17.9 \\
Continuing-gen & 44.1 & 13.5 \\
\midrule
High need & 45.8 & 17.0 \\
Low need & 40.5 & 12.8 \\
\bottomrule
\end{tabular}
\label{tab:group_dropout_rate}
\end{table}

\begin{table}[h]
  \centering
  \caption{Welch two-sample t-test results and Cohen's d effect size of individual ranking change. Positive ranking change means increased predicted dropout risks from BLIND to AWARE model.}~\label{tab:rankingchange}
  \small
  \begin{tabular}{llcccccc}
    \toprule  
    \multicolumn{2}{l}{Group (Avg. ranking change)} & Rank $\Delta$ & Cohen's $d$\\
    \midrule
        \multicolumn{4}{l}{\textit{Online}}\\
    Female (50.5) & Male (-88.4) & 138.9*** & 0.71 \\
    Non-URM (0.5) & URM (-0.9) & 1.4   & 0.01\\
    Continuing-gen (-13.9) & First-gen (18.6) & -32.5*** & 0.16\\
    Low need (104.5) & High need (-60.0) & 164.5***  & 0.87\\
    \midrule
        \multicolumn{4}{l}{\textit{Residential}}\\
    Female (15.9) & Male (-15.1) & 31.0*** & 0.13 \\
    Non-URM  (13.0) & URM (-22.8) & 35.8*** & 0.15\\
    Continuing-gen (27.2) & First-gen (-62.1) & 89.3*** & 0.38\\
    Low need (6.8) & High need (-7.0) & 13.8*** & 0.06\\
    \bottomrule
    \multicolumn{4}{l}{Significance levels: *** p<0.001; ** p<0.005; * p<0.01}
  \end{tabular}
\end{table}

% overall performance and fairness in regard to multiple protected attributes, models, and performance metrics.

% Overall, the analyses provide evidence that the inclusion of protected attributes in college dropout prediction neither impacts model performance nor does it have a big impact on model fairness. If anything, our results show that low-risk students are ranked as marginally higher risk if protected attributes are added to the model. The results hold for residential and online students, for four protected attributes, and for three model performance metrics.

\section{Discussion and Conclusion}

%summary of results (pls add anything important i may have missed)
We set out to answer a simple question: Should protected attributes be included in college dropout prediction models? This study offers a comprehensive empirical examination of how the inclusion of protected attributes affects the overall performance and fairness of a realistic predictive model. We demonstrate this finding across two large samples of residential and online undergraduate students enrolled at one of the largest public universities in the United States. Our findings show that including four important protected attributes (gender, URM, first-generation student, high financial need) does not have any significant effect on three common measures of overall prediction performance when commonly used features (incoming attributes, enrollment information, academic records) are already in the model. Even when used alone without those features, the group indicators defined by the protected attributes are not highly predictive of dropout, although the actual dropout rates are somewhat higher among minoritized groups. In terms of fairness, we find that including protected attributes only leads to a marginal improvement in fairness by assigning dropout risk scores with smaller gaps between minority and majority groups. However, this trend is not sufficiently large to systematically change the final dropout predictions based on the risk scores, and therefore the formal fairness measures are not significantly different between models with and without protected attributes. %Thus, we find no evidence to support the inclusion of protected attributes in college dropout prediction solely on the basis of overall model performance (RQ1).

%Implications for theory and practice
In short, our results suggest limited effects of including protected attributes on the performance of college dropout prediction. This does not point to a clear answer to our normative question and prompts us to further reflect on the focal issue of using protected attributes. Recent work in the broader machine learning community has been in favor of ``fairness through awareness''~\cite{dwork2012}, and has specifically suggested that race-aware models are fairer for student success prediction because they allow the influence of certain features to differ across racial groups~\cite{Kleinberg2018}. Our findings resonate with these existing studies around fairness but only to a marginal extent.
%This applies to features that are subject to group biases; for example, if SAT math scores are systematically lower for women than men, the model could assign different feature weights to these scores for different genders. 
Notably, student groups with historically higher dropout rates are slightly compensated by being ranked lower in predicted dropout risks when protected attributes are used. This compensating effect, however, does not accumulate to statistically significant changes in predicted labels, possibly because the group differences in dropout rates were not sizeable in the past at the institution we study (see Table~\ref{tab:group_dropout_rate}). In other words, protected attributes might have more to contribute to the fairness of prediction in the presence of substantial existing inequalities. 
%Morevoer, even in the absence of pre-processing techniques like orthogonalization, the extent to which other features in the model can be combined to reconstruct the protected attributes is very limited. This finding reaffirms that the inclusion of PAs does not offer notable improvement to model performance.
Still, the existence of a weak compensating instead of segregating effect justifies the inclusion of these attributes. After all, a major argument for race-aware models, and more generally socio-demographic-aware models, is to capture structural inequalities in society that disproportionately expose members of minoritized groups to more adverse conditions.
% The dilemma of whether to include protected attributes stems from the fact that demographics can serve as imperfect proxies for various intrinsic (e.g., psychological) and external (e.g., institutional) factors that affect the success of individual students. If we could perfectly capture all nuances of individual experiences and use them to accurately forecast upcoming challenges for each student, it would not be necessary to take into account the socio-economic groups a student belongs to, as they do not add additional predictive value. In reality, however, we do not know the full spectrum of those variables, but we know that structural inequalities in society disproportionately expose members of minoritized groups to more adverse conditions. In this sense, using protected attributes to predict student outcomes is a feasible solution to capturing factors that on average differ across groups at the cost of ignoring within-group individual differences, i.e., treating each student as a possibly non-existent ``average'' student in the group(s).
In addition, the deliberate exclusion of protected attributes from dropout prediction models can be construed as subscribing to a ``colorblind'' ideology, which has been criticized as a racist approach that serves to maintain the status quo~\cite{burke2018colorblind}.

Another contribution of this work lies in our approach to fairness evaluation. The analyses and visualizations we present are the result of many iterations to arrive at simple yet compelling ways to communicate fairness at different levels of aggregation and across many protected attributes. These methods can be used by those who seek to evaluate model fairness for research and practice. Prior research has mostly focused on evaluating one protected attribute at a time, but in most real-world applications we care about more than one protected attribute. We recommend comparing AWARE against BLIND models in terms of the individual ranking differences by group (Figure~\ref{fig:ranking}) as well as the group difference plots for multiple performance metrics and protected attributes (Figure~\ref{fig:fairness}). This approach offers a sensitive instrument for diagnosing fairness-related issues in various domains of application, which could easily be implemented in a fairness dashboard that evaluates multiple protected attributes, models, and performance metrics~\cite{williamson2021lad}. This will remain a promising line of our future work.

%limitations
%The generalizability of our findings is limited by the specifics of our empirical context. The fact that we see little difference as a result of including protected attributes in our prediction model may not translate to other prediction scenarios with different populations and different prediction targets. However, our general findings are robust to the choice of prediction algorithm (e.g., we also tested additive linear models like logistic regression which gave similar results) and the choice of prediction target (e.g., we also tested more proximal definitions of dropout). Our findings offer an important new perspective into one of the most contentious choices in applied analytics in higher education and we encourage research and practitioners to use model fairness analysis to evaluate the consequences of including protected attributes in their predictive models.
%Conclusion
This research has broader implications for using predictive analytics in higher education beyond its contributions to algorithmic fairness. With a common set of institutional features, we achieve 76\% prediction accuracy and 67\% recall on unseen students in online settings, that is, correctly identifying 67\% of actual dropouts with their first-term records. For residential students, we achieve a higher accuracy of 84\% but a lower recall of 54\%. These performance metrics may seem somewhat lower than in prior studies of dropout prediction, but this might be because most existing studies examine a smaller sample of more homogeneous students, such as students in the same cohort or program~\cite{DelBonifro2020,Dekker2009}. This highlights the general challenge of predicting college dropout accurately. As suggested by the large variance in predicted probabilities for dropouts (Figure~\ref{fig:overall_prob_dist}), widely used institutional features might not perform well in capturing common signals of dropout. This may point to important contextual factors that our institutional practices are presently overlooking. We view this as a limitation and important next step that will require both an interrogation of the theoretical basis for predictors and close collaboration with practitioners.
%Stakeholders who feel conflicted about including or excluding protected attributes like gender or race can rest assured that it may not make any difference at all in terms of accuracy or fairness.

Further directions for future research in this area include exploring counterfactual notions of fairness in this context by testing how predictions would differ for counterfactual protected attributes, all else being equal. This would benefit the contemporary education system which relies increasingly on research that provides causal evidence. We would also like to move from auditing to problem-solving by evaluating correction methods for any pre-existing unfairness in predictions to see how the AWARE relative to the BLIND model responds~\cite{lee2020evaluation}. We hope that this study inspires more researchers in the learning analytics and educational data mining communities to engage with issues of algorithmic bias and fairness in the models and systems they develop and evaluate.

% \begin{table}[]
%   \caption{Subgroup Performance of AWARE vs. BLIND}
%   \label{tab:subgroup_performancee}
% \resizebox{\textwidth}{!}{%
% \begin{tabular}{lc|ccc|ccc|ccc|ccc}
% \toprule
%  & \multicolumn{1}{l|}{} & Male & Female & \Delta & URM & \begin{tabular}[c]{@{}c@{}}Non-\\ URM\end{tabular} & \Delta & FirstGen & \begin{tabular}[c]{@{}c@{}}Non-\\ FirstGen\end{tabular} & \Delta & HighNeed & \begin{tabular}[c]{@{}c@{}}Non-\\ HighNeed\end{tabular} & \Delta \\ 
%  \midrule
% \multicolumn{1}{c|}{Accuracy (\%)} & AWARE & 74.6 & 76.8 & -2.2** & 77.3 & 75.3 & 2* & 76.6 & 75.6 & 1 & 76.3 & 75.4 & 0.9 \\
% \multicolumn{1}{l|}{} & BLIND & 74.3 & 76.5 & -2.2** & 77 & 75 & 2* & 76.4 & 75.2 & 1.2 & 76.2 & 74.9 & 1.3 \\ 
% \midrule
% \multicolumn{1}{c|}{Recall (\%)} & AWARE & 64.1 & 63.6 & 0.5 & 70.2 & 60 & 10.2*** & 64.5 & 63.3 & 1.2 & 65.2 & 61 & 4.2** \\
% \multicolumn{1}{l|}{} & BLIND & 63.5 & 63.7 & -0.2 & 70.1 & 59.7 & 10.4*** & 65.1 & 62.5 & 2.6 & 65 & 60.8 & 4.2** \\ 
% \midrule
% \multicolumn{1}{c|}{FPR (\%)} & AWARE & 16.2 & 14.9 & 1.3 & 17.1 & 14.5 & 2.6** & 15 & 15.7 & -0.7 & 15.2 & 15.6 & -0.4 \\
% \multicolumn{1}{l|}{} & BLIND & 16.2 & 15.5 & 0.7 & 17.5 & 14.8 & 2.7** & 15.6 & 15.8 & -0.2 & 15.3 & 16.3 & -1\\
% \bottomrule
% \multicolumn{}{}{}
% \end{tabular}%
% }
% \end{table}

% \begin{acks}
% Left blank.
% \end{acks}
% BALANCE COLUMNS
\balance{}

% REFERENCES FORMAT
% References must be the same font size as other body text.
\bibliographystyle{SIGCHI-Reference-Format}
\bibliography{sample}

\end{document}